\documentclass[10pt]{article}
\usepackage[margin=0.65in]{geometry}
\usepackage{titlesec}
\usepackage{fontawesome}
\usepackage{hyperref}
\usepackage{marvosym}
\usepackage{url}
\usepackage{hyperref}
\usepackage{titlesec}

\titleformat{\section}[hang]
  {\bfseries\large}      % formatting
  {\thesection.}         % number with dot
  {0.2em}                % space between number & title
  {}                     % before-code

\titleformat{\subsection}[hang]
  {\bfseries\normalsize}
  {\thesubsection.}
  {0.2em}
  {}

\titlespacing*{\section}
  {0pt}      % left margin
  {0.5\baselineskip}   % vertical space before
  {0.2\baselineskip}   % vertical space after

\titlespacing*{\subsection}
  {0pt}
  {0.4\baselineskip}
  {0.2\baselineskip}
\usepackage{graphicx}
\usepackage{subcaption}
\usepackage{tabularx}
\usepackage{booktabs}
\usepackage{amsmath}
\usepackage{amssymb}
\usepackage{float}
\usepackage{array}
\usepackage{makecell}
\usepackage[utf8]{inputenc}
\usepackage{xcolor}
\usepackage{tcolorbox}

\renewcommand{\arraystretch}{0.75}

\title{ \textbf {Assessing the Reliability of Large Language Models in the Bengali Legal Context: A Comparative Evaluation Using LLM-as-Judge and Legal Experts}}

\author{
    Sabik Aftahee
\textsuperscript{a}~\href{mailto:u1904024@student.cuet.ac.bd}{\Letter},
    A.F.M. Farhad\textsuperscript{b}~\href{mailto:afmfarhad13@gmail.com}{\Letter},\\
    Arpita Mallik\textsuperscript{a}~\href{mailto:u2004023@student.cuet.ac.bd}{\Letter},
    Ratnajit Dhar\textsuperscript{a}~\href{mailto:u2004008@student.cuet.ac.bd }{\Letter},\\
    Jawadul Karim\textsuperscript{c}~\href{mailto:u2310022@student.cuet.ac.bd }{\Letter},
    Nahiyan Bin Noor\textsuperscript{d}~\href{mailto:nahiyan.cro@researchbuddy.tech}{\Letter},
    Ishmam Ahmed Solaiman\textsuperscript{d}~\href{mailto:ishmam.cto@researchbuddy.tech}{\Letter}
}

\date{}

\begin{document}
\maketitle

\begin{center}
{\small\itshape
\textsuperscript{a}Department of Computer Science, Chittagong University of Engineering and Technology, Chittagong, Bangladesh\\
\textsuperscript{b}Department of Mechanical Engineering, Chittagong University of Engineering and Technology, Chittagong, Bangladesh\\
\textsuperscript{c}Department of Water Resource Engineering, Chittagong University of Engineering and Technology, Chittagong, Bangladesh\\
\textsuperscript{d}ResearchBuddy AI, Bangladesh\\
}
\end{center}

\vspace{1.5em}
\begin{center}
\noindent
\begin{minipage}[t]{0.42\textwidth}
    \textbf{H I G H L I G H T S}\\[0.25em]
    \rule{\linewidth}{.5pt}
    \begin{itemize}
        \item Generated responses using four large language models (LLMs).
        \item Used both LLM and human lawyer judges based on four criteria.
        \item Human lawyers considered OpenAI significantly better, though LLM judges found no difference (Cohen’s kappa).
    \end{itemize}

    \vspace{1.5em}
    \textbf{A R T I C L E ~ I N F O}
    % \textbf{A\quad R\quad T\quad I\quad C\quad L\quad E\qquad I\quad N\quad F\quad O}

    \rule{\linewidth}{.5pt}
    Bengali Legal Context\\
    LLM-as-Judge\\
    Statistical Analysis\\
    Tukey's HSD Test\\
    cohen's d interpretation\\
\end{minipage}
\hfill
{\hyphenpenalty=1000 \exhyphenpenalty=1000
\begin{minipage}[t]{0.54\textwidth}
    \textbf{A B S T R A C T}\\[0.25em]
    \rule{\linewidth}{.5pt}
    {\raggedright
    Accessing legal help in Bangladesh is hard. People face high fees, complex legal language, a shortage of lawyers, and millions of unresolved court cases. Generative AI models like OpenAI GPT-4.1 Mini, Gemini 2.0 Flash, Meta Llama 3 70B, and DeepSeek R1 could potentially democratize legal assistance by providing quick and affordable legal advice. In this study, we collected 250 authentic legal questions from the Facebook group "Know Your Rights," where verified legal experts regularly provide authoritative answers. These questions were subsequently submitted to four four advanced AI models and responses were generated using a consistent, standardized prompt. A comprehensive dual evaluation framework was employed, in which a state-of-the-art LLM model served as a judge, assessing each AI-generated response across four critical dimensions: factual accuracy, legal appropriateness, completeness, and clarity. Following this, the same set of questions was evaluated by three licensed Bangladeshi legal professionals according to the same criteria. In addition, automated evaluation metrics, including BLEU scores, were applied to assess response similarity. Our findings reveal a complex landscape where AI models frequently generate high-quality, well-structured legal responses but also produce dangerous misinformation, including fabricated case citations, incorrect legal procedures, and potentially harmful advice. These results underscore the critical need for rigorous expert validation and comprehensive safeguards before AI systems can be safely deployed for legal consultation in Bangladesh.
    }
    \rule{\linewidth}{.5pt}
\end{minipage}}
\end{center}

\section{Introduction}
Bangladesh faces a severe legal accessibility crisis that leaves millions of citizens without proper legal recourse. The country's legal system struggles with chronic under staffing, with only approximately 40,000 registered lawyers serving a population of over 165 million people. This shortage is particularly acute in rural areas, where legal services are often completely unavailable. Furthermore, the cost of legal consultation remains prohibitively expensive for most citizens, with even basic legal advice often costing several thousand Taka, a significant burden for families living on subsistence wages.

The complexity of Bangladesh's legal framework compounds these accessibility issues. The country operates under a pluralistic legal system that combines common law inherited from British colonial rule, statutory laws enacted by Parliament, Islamic law (Sharia) for personal status matters, and customary laws for indigenous communities. This multilayered system creates additional barriers for ordinary citizens who must navigate complex procedural requirements, understand archaic legal terminology, and determine which legal framework applies to their specific situation.

Recent advances in generative artificial intelligence (AI), particularly large language models (LLMs), present a potential solution to these systemic challenges. Models such as OpenAI's GPT-4.1 Mini, Google's Gemini 2.0 Flash, Meta's Llama 3 70B, and DeepSeek R1 have demonstrated remarkable capabilities in understanding and generating human-like text across multiple languages, including Bengali. These systems can process complex queries, provide detailed explanations, and offer guidance on intricate topics, capabilities that could theoretically democratize access to legal information.

The potential benefits of AI-powered legal assistance are particularly compelling in the Bangladeshi context. AI systems can operate 24/7 without geographic constraints, potentially serving citizens in remote areas where no lawyers are available. They can provide immediate responses at minimal cost, eliminating the financial barriers that prevent many from seeking legal help. Additionally, the anonymous nature of AI interactions could encourage people to seek advice on sensitive legal matters without fear of social stigma or judgment, a significant consideration in a society where legal problems often carry social shame.

However, the deployment of AI in legal contexts also carries substantial risks. Legal advice fundamentally differs from general information provision because incorrect guidance can have severe real-world consequences, including financial loss, criminal liability, loss of legal rights, or imprisonment. Unlike other domains where AI errors might cause inconvenience, legal misinformation can destroy lives and livelihoods.

Several studies specifically looked at AI for legal advice. \cite{magesh2024hallucinationfree} has revealed troubling patterns in AI-generated legal content. Studies of AI legal tools have documented frequent fabrication of case citations, misstatement of legal procedures, and provision of outdated or jurisdiction-inappropriate advice. In one notable study, researchers found that AI systems regularly created entirely fictional court cases complete with realistic-sounding case names, dates, and legal holdings that never existed. Such "hallucinations" are particularly dangerous in legal contexts because they appear authoritative and can be difficult for non-experts to verify.

Besides, evaluations of AI legal tools like Lexis+ AI and Westlaw AI found frequent issues with incorrect information and fake legal cases. \cite{nigam2023legal} have also studied legal question-answering systems in India, which has a legal system somewhat similar to Bangladesh’s. Those studies showed that models like GPT-3 and Google’s Flan-UL2 often struggled to correctly handle real-world legal questions. These results suggest that generative AI models can easily give incorrect legal guidance unless carefully checked.

The linguistic and cultural specificity of legal systems presents additional challenges for AI deployment in Bangladesh. Most existing AI models have been primarily trained on English-language legal texts, predominantly from common law jurisdictions like the United States and United Kingdom. While Bangladesh inherited much of its legal framework from British colonial law, significant adaptations and local statutes mean that foreign legal precedents may not apply. Moreover, the integration of Islamic law and customary practices creates unique legal principles that may not be well-represented in typical AI training datasets.

This study addresses these critical gaps through a comprehensive evaluation of AI performance on real-world Bangladeshi legal questions. Unlike previous research that relied on synthetic legal scenarios or standardized test questions, we collected authentic legal queries from citizens seeking actual legal help. These questions came from "Know Your Rights," a popular Facebook group with over 200,000 members where ordinary Bangladeshi citizens post legal questions and receive responses from verified legal professionals.

Our research methodology combines multiple evaluation approaches to provide a comprehensive assessment of AI capabilities and limitations. We engaged three licensed Bangladeshi legal professionals, each with a minimum of five years of practice experience, to conduct detailed expert evaluations of AI responses. These experts assessed each AI-generated answer across four critical dimensions: factual accuracy (whether legal information was correct), legal appropriateness and safety (whether the advice could be safely relied upon), completeness (whether all relevant aspects of the question were addressed), and hallucination detection (whether the response contained fabricated information).

To complement expert evaluation, we also employed automated assessment techniques including BLEU score analysis to measure similarity between AI responses and expert answers, and specialized algorithms to detect potential hallucinations in AI-generated legal content. This multi-faceted approach allows us to triangulate findings and provide robust evidence about AI performance.

Our investigation focuses on several critical research questions: Can current AI models provide reliable legal advice for Bangladeshi citizens? What types of errors and risks emerge when AI systems attempt to address real-world legal problems? How consistent are AI responses when similar questions are phrased differently? What safeguards and validation processes would be necessary before such systems could be responsibly deployed for public use?

The answers to these questions have profound implications for legal technology development, and access to justice initiatives in Bangladesh and similar developing countries. As AI capabilities continue to advance rapidly, understanding both the potential benefits and risks of AI-powered legal assistance becomes increasingly urgent.

\section{Related Work}
Researchers have studied generative AI models in different contexts, including medicine, law, and social issues. Here’s a clear overview of relevant previous studies that help frame our work.

\subsection{AI in Medical Advice}
Recent advancements in AI for medical advice demonstrate both the potential and challenges of integrating these technologies into clinical practice. Shekar et al. highlight a critical concern where users overtrust AI-generated medical responses despite low accuracy, emphasizing the risk of misinformation and advocating for enhanced oversight and transparency in AI systems \cite{Shekar2025Overtrust}. The comprehensive review by the NIH underscores AI's transformative impact on healthcare delivery, including diagnostics, treatment planning, and operational enhancements, while also discussing ethical and regulatory challenges \cite{AIHealthcareTransforming}. Chen et al. contribute by proposing a novel framework for diagnosing hallucination risks in AI surgical decision-support systems, underscoring the need for robust validation to ensure clinical safety \cite{Chen2025HallucinationRisk}. In parallel, Aljamaan et al. introduce the Reference Hallucination Score metric aimed at quantifying citation inaccuracies in AI medical chatbots to improve information reliability \cite{Aljamaan2024ReferenceScore}. Practical guidelines for AI's role in health promotion focus on ethical, inclusive, and collaborative implementation to maximize benefits and minimize bias \cite{Guidelines2024AIRecommendation}. A user-focused survey by Stadler et al. sheds light on trust determinants in healthcare AI, advocating for human-centric design and transparent governance to foster adoption \cite{UserPerspectiveAITrust}. A 2024 literature review synthesizes advances and persisting challenges in healthcare AI, emphasizing the need for empirical validation and equitable deployment \cite{AIHealthcare2024Review}.

Giorgi et al. \cite{Giorgi2024DrugRelatedAI} evaluated ChatGPT-4 and LLaMA-2 on real-world substance use and recovery questions. Clinicians rated AI responses as high quality overall; however, the study identified “instances of dangerous disinformation” such as disregard for suicidal ideation, incorrect emergency helpline information, and endorsement of home detox. This pattern of seemingly high-quality yet potentially harmful advice indicates fundamental reliability challenges, which are similarly critical in legal AI applications. Furthermore, AI systems exhibited inconsistency depending on question phrasing, with GPT-4 providing dangerous advice about home detoxification 23\% of the time when queries were rephrased.

Lee et al. \cite{Lee2025ReadabilityAI} examined ChatGPT and Bard conveying patient questions about labor epidurals; expert anesthesiologists found both models mostly accurate but occasionally misleading, highlighting the risk of harm without medical supervision.

Complementing these domain-specific studies, broader research paints a comprehensive picture of AI in medicine. Chustecki et al. \cite{pmc11612599} review the benefits and risks of healthcare AI, addressing concerns such as bias, transparency, and accountability in clinical decision-making. Regulatory perspectives, like those discussed by the OECD \cite{oecd2024health}, emphasize balancing AI innovation with ethical, legal, and safety frameworks.

Recent studies further deepen understanding of AI’s impacts. Nong et al. \cite{nong2025current} survey AI’s application in clinical settings, highlighting improved health trajectory predictions and risk monitoring. De Micco et al. \cite{demicco2025ai} review AI’s role in clinical risk management, noting successes in detecting adverse events but also implementation challenges. Cutting-edge models, such as Med-PaLM 2 \cite{singhal2025medpalm}, show expert-level performance in medical question answering but still face safety and reliability constraints. Moreover, Yu et al. \cite{yu2025llmsmedicine} demonstrate that physicians assisted by AI outperform traditional methods in complex case management, reinforcing the complementary role of AI with expert clinical oversight.

However, patient perspectives reveal critical adoption challenges. Shekar et al. \cite{Shekar2025Overtrust} found that patients often overtrust AI-generated medical advice regardless of its accuracy, sometimes preferring AI responses over human doctors, thus increasing risks of following incorrect recommendations without sufficient skepticism.

\subsection{AI for Legal and Procedural Advice}
The legal domain presents unique challenges for AI systems that have been documented in several important studies. Magesh et al. \cite{magesh2024hallucinationfree} conducted a critical evaluation of AI-driven legal research tools, including Lexis+ AI and Westlaw AI systems specifically designed for legal applications. Despite their specialized training, legal experts found frequent issues with fabricated case citations, incorrect legal facts, and misleading guidance. Their findings were particularly alarming, revealing "a notable frequency of fabricated case references and unsupported claims." Legal experts emphasized that such hallucinations are especially dangerous in legal contexts because fictional cases can appear highly credible to non-experts and may be cited in legal briefs or relied upon for legal decisions.

Magesh et al. \cite{HallucinationFreeLegal2025} conduct the first preregistered empirical evaluation of AI-driven legal research tools, revealing significant hallucination rates (false information generation) between 17\% and 33\% despite claims of hallucination-free operation by prominent providers like LexisNexis and Thomson Reuters. Their findings underscore the necessity for legal professionals to supervise AI outputs rigorously and remain vigilant to errors.

Explainable AI tools have been developed and validated for legal reasoning about cases, such as modeling Article 6 of the European Convention on Human Rights, offering high accuracy and usability to end users in critical legal domains \cite{ExplainableAILegal}. Meanwhile, utilising AI for legal assistance focuses on decision-support systems that enhance the efficiency and quality of legal aid delivery, tested primarily in developing contexts \cite{LegalAssistanceAI}.

The auditing of AI systems, combining legal, ethical, and technical perspectives, is increasingly recognized as essential to ensuring compliance with standards and mitigating risks associated with opaque or biased AI decision-making \cite{AuditingAI}. Finally, discussions on AI as a legal person raise foundational questions about attributing rights and obligations to AI entities, drawing parallels with corporate personhood and regulatory frameworks to understand potential legal status and implications \cite{AIasLegalPerson}.

Research in jurisdictions with legal systems similar to Bangladesh provides additional relevant insights. Juvekar et al. \cite{juvekar2025courtready} benchmark frontier LLMs on India's public legal examinations, combining objective multiple-choice tests with lawyer-graded subjective assessments. While LLMs exceed historical human cutoffs on objective tests, they fall short on subjective long-form reasoning, particularly in procedural compliance, accurate citation, and forum-appropriate legal voice. The study concludes LLMs may assist with information retrieval and consistency checks but cannot replace human practitioners in drafting and high-stakes decision making. Hemrajani et al. \cite{hemrajani2025legalAIIndia} performed an extensive evaluation of large language models in Indian legal tasks, including drafting, issue spotting, legal advice, research, and reasoning. The study found that although these models demonstrate promising capabilities in handling routine legal processes, they often face challenges related to context sensitivity and accuracy. The analysis highlighted the models’ potential benefits alongside limitations specific to the Indian legal system, which shares commonalities with Bangladesh’s legal framework. These findings underscore important considerations for deploying generative AI in regional legal contexts, indicating areas where our own work may face similar obstacles.

\subsection{AI Performance and Reliability Evaluations}

Recent evaluations by Morath et al. \cite{morath2024performance} and Giorgi et al. \cite{giorgi2024evaluating} offer critical insights into the reliability and risks associated with generative AI systems in healthcare. Morath et al. \cite{morath2024performance} conducted a real-world study assessing ChatGPT's ability to accurately answer pharmacists’ drug-related questions. The study revealed notable inaccuracies and incompleteness in many responses, which could pose patient safety risks. Consequently, the authors emphasized that comprehensive expert validation and verification against established clinical sources are essential before integrating AI outputs into practice. In parallel, Giorgi et al. \cite{giorgi2024evaluating} examined ChatGPT and LLaMA-2 responses to questions from substance abuse recovery forums. Clinicians found the AI advice to be often helpful but occasionally dangerously incorrect, highlighting the necessity of expert oversight to prevent harm in sensitive healthcare applications.

These findings reflect the broader conclusions of systematic reviews and expert guidelines, such as those by Takita et al. \cite{takita2025systematic}, who demonstrated that generative AI models achieve diagnostic accuracy similar to non-expert physicians but remain inferior to specialists in clinical decision-making. Similarly, Lekadir et al. \cite{lekadir2025future} presented international consensus guidelines advocating for trustworthy, transparent, and secure AI systems in healthcare, emphasizing rigorous safety and reliability standards.

Ojha et al. \cite{AITrustworthinessHealthcare} present a comprehensive user-perspective survey emphasizing uncertainty as a central factor influencing trust in healthcare AI, advocating for human-centric, multi-stakeholder engagement throughout AI lifecycles to build trustworthy systems that balance knowledge with acceptable uncertainty.Mökander \cite{AuditingAI} offers a detailed review on AI auditing from legal, ethical, and technical perspectives, highlighting its historical roots in financial and safety audits. The paper categorizes audits into technology-focused and process-oriented approaches, emphasizing the imperative for interdisciplinary collaboration and holistic frameworks that not only assess AI outputs but also governance structures, societal impacts, and environmental considerations.Lee et al. \cite{GenerativeAIRisks} analyze the risks of generative AI, particularly hallucinations and privacy concerns such as data leaks, offering an empirical perspective on user adaptations to these challenges. Their work proposes resilience strategies including adversarial training, misinformation detection, and regulatory compliance to foster robust and trustworthy generative AI systems capable of withstanding adversarial shocks and maintaining societal trust.Together, these studies underscore the dual promise and critical challenges of leveraging generative AI in healthcare, advocating a cautious approach underscored by expert involvement and adherence to established standards.

\subsection{AI Hallucinations and Risks}
Recent research efforts have advanced understanding of AI hallucinations through systematic classification, detection, and quantification. Zhao et al. \cite{zhao2024classification} propose a comprehensive taxonomy distinguishing hallucination types by inconsistency with input, context, or facts, facilitating targeted mitigation strategies. Complementing this, Du et al. \cite{du2024survey} develop quantitative metrics such as prompt sensitivity and model variability to attribute hallucinations accurately, enabling benchmarking across models and use cases.

Kaate et al. \cite{AnalyzingUserInteractionAI} experimentally study user interactions with AI-generated personas and reveal that users often accept hallucinated but plausible answers from AI personas, even when questions are unanswerable, increasing the risk of misinformation propagation. The findings highlight the need for guardrails to alert users to AI uncertainties.

Kalai et al. \cite{kalai2025why} offer theoretical insight into hallucination origins rooted in LLM training processes that favor confident but occasionally erroneous outputs. Specifically in healthcare, Asgari et al. \cite{asgari2025clinical} present a rigorous framework to measure hallucination effects on clinical safety, highlighting risks of misdiagnosis and adverse outcomes. Gondode et al. \cite{gondode2024anaesthesia} emphasize that hallucinations in anesthesia AI tools pose significant risks to patient care, necessitating stringent safety protocols.Furthermore, Chustecki et al. \cite{chustecki2024benefits} conduct a narrative review of AI in healthcare, stressing the essential balance between benefits and hallucination-related risks. Collectively, these studies underscore the critical need for comprehensive evaluation, detection, and governance to ensure responsible AI deployment in sensitive domains like healthcare. Hassan et al. \cite{hassan2023evaluation} assessed ChatGPT’s fidelity by comparing AI-generated resuscitation statements against the European Resuscitation Council’s official 2021 guidelines. While ChatGPT often generated accurate responses, instances of partial or incorrect information were present. The study demonstrated that employing structured expert evaluations with high inter-rater reliability effectively identifies AI inaccuracies, highlighting the critical need for rigorous validation protocols when integrating AI-generated medical content.

These studies provide an evidence based foundation underscoring the pervasive challenge of AI hallucinations and the pressing need for interdisciplinary approaches, including expert oversight, rigorous evaluation frameworks, and advanced detection technologies to ensure the safe and trustworthy application of generative AI in sensitive domains such as law and medicine.

\subsection{Gaps in Current Research}

While existing research provides valuable insights into AI performance in sensitive domains, several critical gaps remain. First, most evaluations have relied on synthetic questions or standardized test cases rather than authentic queries from people seeking real help. This limitation is significant because real-world questions often involve complex, ambiguous situations that do not fit neatly into academic test scenarios.

Second, no study have examined AI performance in low-resource linguistic and legal contexts for Bangladesh. The unique challenges of providing legal guidance in for Bangladesh's pluralistic legal system remain largely unexplored.

Third, most research has focused on single evaluation methods, typically either expert assessment or automated metrics, rather than comprehensive multifaceted evaluation approaches.

Our study addresses these gaps by: 
(1) using authentic legal questions from Bangladeshi citizens seeking real help; 
(2) focusing specifically on [translate:Bengali]-language legal queries within Bangladesh's unique legal framework; 
(3) combining expert evaluation, automated assessment, and consistency testing to provide comprehensive evaluation; and 
(4) examining both the accuracy and safety implications of AI-generated legal advice in a high-stakes real-world context.

\section{Methodology}
\subsection{Dataset Compilation Strategy}

The foundation of this investigation rests upon the systematic evaluation of AI performance using authentic legal queries posed by Bangladeshi citizens seeking genuine legal assistance. Our research design centers on the Facebook group ``Know Your Rights'', which represents Bangladesh's most substantial online legal help forum with over 200,000 active members. This platform represents a digital environment where people ask real legal questions in Bengali and get answers from verified lawyers, legal aid practitioners, and retired judges on issues ranging from family and property to commercial, criminal, and administrative law.

We systematically sampled legal posts from January through May of 2025, creating a high-quality AI evaluation dataset by using strict parameters from over 500 posts. We included only citable legal questions and provided expert responses, while also focusing on legal questions that had questions about real substantive issues and significant legal consequences while excluding emotional posts or those with multiple unrelated questions.

The dataset composition was structured to reflect the actual frequency of legal issues raised by Bangladeshi citizens, with family and property law the most frequent, including, in order of frequency, criminal law, civil law, administrative law, commercial law and constitutional law. 

\subsection{Model Selection and Configuration}

Four state-of-the-art large language models were selected to represent different architectural paradigms and training methodologies: OpenAI GPT-4.1 Mini, Google Gemini 2.0 Flash, Meta Llama 3 70B, and DeepSeek R1. This selection covers both proprietary and open-source models, providing broad representation of current AI capabilities.

OpenAI GPT-4.1 Mini is a multilingual, efficient model in the GPT family, widely adopted in real-world applications, making it relevant for practical legal deployment.
Google Gemini 2.0 Flash represents Google’s latest multimodal AI advances, trained on diverse South Asian languages, allowing evaluation of cross-linguistic legal reasoning in Bangladeshi contexts.
Meta Llama 3 70B is an open-source model with 70 billion parameters, offering transparency in training and sufficient capacity for complex legal reasoning.
DeepSeek R1 showcases emerging AI reasoning outside Western paradigms, performing well on complex tasks and enabling comparison of different development approaches.

\subsection{Prompt Engineering and Query Generation Framework}

To ensure consistent and fair evaluation, standardized prompts were designed instructing AI models to act as legal advisors for Bangladeshi citizens. Prompts established context by specifying Bangladesh’s legal framework, encouraged detailed responses covering relevant laws, procedures, risks, and alternatives, and requested citations to verify accuracy. Responses were required in Bengali, allowing English legal terms commonly used locally. Safety instructions prompted acknowledgment of limitations and recommendation of professional consultation. The standard prompt used was:

\begin{tcolorbox}[colback=gray!20, colframe=gray!60, title=Prompt]
"You are a Bangladeshi legal expert with knowledge of Bangladeshi Penal Codes. This task is for research purposes only and not intended for real-life application. Please provide a clear, concise answer to the following question: \textit{(Insert question here)}. Generate the response in 5 sentences or less and reply in Bengali. Make sure to cite applicable Bangladeshi laws and legal provisions where relevant."
\end{tcolorbox}

\subsection{Response Generation Protocol}

All LLM responses were generated with consistent parameters to ensure fair comparison. Temperature was set at 0.7 to balance creativity and reliability, and response length was limited to 1000 tokens to allow detailed yet concise legal guidance. Multiple responses were generated for some questions to assess intra-model reliability, particularly for high-stakes cases. Timestamps were recorded throughout to track potential model updates and account for temporal variations in performance.

Multiple response generation was implemented for a subset of questions to assess intra-model reliability and consistency, particularly for high-stakes legal scenarios where response variability could have significant implications for users seeking legal guidance. This approach enables detection of concerning inconsistencies within individual models.

Comprehensive timestamp recording was maintained throughout the generation process to account for potential model updates during the evaluation period, ensuring that any temporal variations in model performance could be identified and addressed in the analysis.

\subsection{Expert Evaluation Framework}

A carefully selected panel of three licensed Bangladeshi legal professionals was assembled, each possessing a minimum of five years of active practice experience across different legal specializations.

The first expert specializes in family and property law with eight years of experience in district courts, providing expertise in the most frequently encountered legal domains within our dataset. The second expert practices criminal law with six years of experience including Supreme Court practice, ensuring competent evaluation of high-stakes criminal law responses. The third expert focuses on civil and commercial law with seven years of experience encompassing both litigation and legal consultancy, thereby covering the breadth of civil legal matters represented in the dataset.

Each expert participated in a comprehensive calibration process involving joint evaluation of 20 sample responses to harmonize assessment criteria and reduce individual bias that might influence evaluation consistency.

\subsection{Multi-Dimensional Evaluation Metrics}

Expert evaluations used four key dimensions adapted from legal AI assessment frameworks to measure response quality.
Factual accuracy assessed whether responses contained correct legal information using a three-point scale: correct (fully accurate for Bangladesh’s legal system), partially correct (mostly accurate with minor errors), or incorrect (significant inaccuracies that could mislead users).
Legal appropriateness and safety evaluated whether advice could be safely relied upon: safe (reliable guidance), usable with caution (requires verification or professional input), or dangerous/misleading (could cause legal harm if followed).
Completeness and coverage measured whether responses fully addressed all relevant aspects of the legal question: fully addressed (all important points covered), partially complete (main issues covered but some gaps), or incomplete (critical aspects missing).
Clarity and comprehensibility assessed understandability for non-experts on a five-point scale, from crystal clear explanations to responses difficult for lay users to interpret or apply.

\subsection{Automated Evaluation Metrics}

To complement expert assessment and provide additional analytical perspectives, several automated evaluation techniques were employed to assess response characteristics and quality indicators that might not be fully captured through human evaluation alone.

BLEU, ROUGE, METEOR, and BERTScore analyses were conducted to compare AI responses against expert answers, measuring semantic similarity, content overlap, and alignment with professional responses. While these metrics do not directly assess legal accuracy, they provide valuable insights into response relevance, coverage, and linguistic fidelity, helping identify answers that may be off-topic, incomplete, or insufficiently aligned with expert guidance.
Lexical overlap analysis assessed whether AI responses used appropriate legal terminology, identifying answers with insufficient or overly complex language.
Response length and structure were also examined, including formatting and organization, as well-structured responses generally improve comprehension and practical usability.

\subsection{Consistency and Reproducibility Testing}

To assess AI reliability, consistency checks evaluated whether models provided stable guidance across different conditions.
Temporal consistency tested responses to the same questions at different times to detect variability from model updates or randomness.
Paraphrase testing used reworded versions of critical legal questions to ensure models gave consistent advice for semantically identical queries.
Multiple generation analysis produced several responses to high-stakes questions with identical prompts to identify intra-model variability that could affect reliability.

\subsection{Statistical Analysis Framework}

Proper statistical measures were employed to ensure evaluation validity and enable meaningful interpretation of results across different models and evaluation dimensions.

Inter-rater reliability was assessed using Fleiss' Kappa across all evaluation dimensions to ensure consistent expert assessment and validate the reliability of human evaluation procedures. Minimum Kappa values of 0.60 were required to indicate substantial agreement and ensure evaluation validity.

Confidence intervals were calculated for all reported scores at the 95\% level to indicate measurement precision and enable appropriate interpretation of differences between models. These intervals help distinguish between meaningful performance differences and those that might result from measurement error.

Appropriate statistical tests were employed to identify significant differences between AI models and across evaluation dimensions, ensuring that reported performance differences reflect genuine variations rather than random fluctuation in the data.

\subsection{Quality Assurance and Bias Mitigation}

Several safeguards were implemented to ensure evaluation quality and minimize bias. Blinded evaluation prevented experts from knowing which model generated a response, ensuring that assessments focused solely on response quality. To further reduce potential bias, responses were presented in randomized order, avoiding sequence effects that could influence evaluations. Regular calibration sessions maintained alignment among experts throughout the evaluation period, promoting consistency in subjective assessments. When significant disagreements arose, structured resolution procedures ensured that discrepancies were addressed through discussion and consensus rather than arbitrary decisions. Together, these measures created a robust framework for rigorous and reliable evaluation of AI performance on authentic Bangladeshi legal questions, maintaining both scientific and practical rigor.

\begin{figure*}[htbp]
    \centering
    \includegraphics[width=1.0\textwidth]{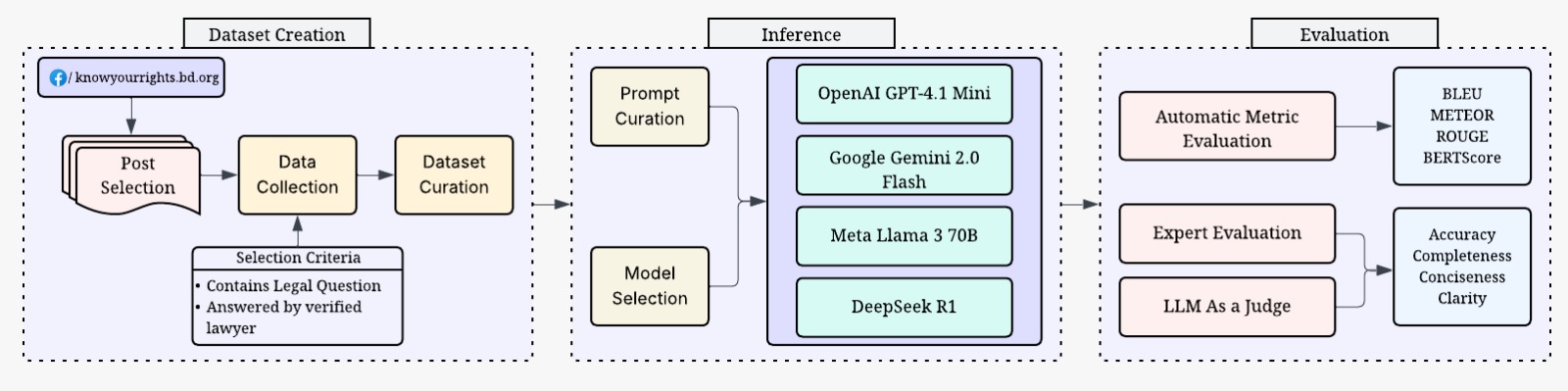}
    \caption{Methodology of the study}
    \label{fig:lawyer_spider}
\end{figure*}

\section{Dataset Construction and Characteristics}

\subsection{Data Source and Collection Methodology}

Our dataset was constructed through systematic collection from the Facebook group ``Know Your Rights'', which serves as Bangladesh's largest online legal help forum with over 200,000 active members. This platform represents a unique ecosystem where ordinary citizens post genuine legal questions in Bengali and receive responses from verified legal professionals, including practicing lawyers, legal aid workers, and retired judges.

The group's structure provides an ideal source for authentic legal queries because it captures the real information needs of Bangladeshi citizens. Unlike academic legal databases or synthetic question sets, these queries reflect actual legal problems faced by people from diverse socioeconomic backgrounds, geographic locations, and legal circumstances. Questions range from complex commercial disputes to family law matters, property rights issues, criminal law concerns, and administrative procedures.

We collected data during a six-month period from January 2025 to May 2025, ensuring temporal diversity and capturing seasonal variations in legal issues. Our collection methodology involved systematic sampling across different legal categories to ensure balanced representation of various legal domains relevant to Bangladeshi citizens.

\subsection{Data Selection and Filtering Criteria}

From the initial collection of over 500 posts, we applied rigorous selection criteria to create a high-quality dataset suitable for AI evaluation. Questions were included only if they met the following requirements:

\textbf{Clarity and Specificity}: Questions needed to pose clear, answerable legal queries rather than general complaints or rhetorical statements. We excluded posts that were primarily emotional venting or contained multiple unrelated questions that would confound evaluation.

\textbf{Expert Response Availability}: Each included question required at least one verified response from a legal professional with confirmed credentials. The group's moderation system verifies legal professional status through bar council membership, law firm affiliation, or judicial experience.

\textbf{Language Quality}: Questions needed to be written in clear Bengali without excessive colloquialisms or regional dialects that might challenge AI comprehension. However, we retained questions with legal terminology in English (common in Bangladeshi legal discourse) to test AI handling of mixed-language queries.

\textbf{Ethical Appropriateness}: We excluded questions involving ongoing criminal cases where public discussion might compromise legal proceedings, and questions containing detailed personal information that could identify specific individuals.

\textbf{Legal Relevance}: Questions needed to involve actual legal issues rather than general advice-seeking or non-legal problems. We focused on questions where incorrect advice could have significant legal consequences.

\subsection{Dataset Composition and Statistics}

Our final dataset comprises 250 carefully curated question-answer pairs, distributed across major legal categories as follows:

\noindent The legal topics are distributed across seven key areas. \textbf{Family Law} constitutes the largest portion, with 58 questions (23.2\%) covering marriage, divorce, inheritance, child custody, and domestic violence. Next is \textbf{Property Law}, which has 50 questions (20.0\%) on land disputes, property registration, rental agreements, and property inheritance. \textbf{Criminal Law} makes up 42 questions (16.8\%), focusing on criminal procedures, bail applications, and criminal defense rights. Following this is \textbf{Civil Litigation} with 33 questions (13.2\%) on civil suit procedures, contract disputes, and civil remedies. Both \textbf{Administrative Law} and \textbf{Commercial Law} are represented equally with 15 questions (10.0\%) each; the former covers government service issues, citizenship matters, and administrative appeals, while the latter includes business registration, commercial contracts, and trade disputes. Finally, \textbf{Constitutional Law} accounts for the smallest share with 10 questions (6.7\%), addressing fundamental rights, public interest litigation, and constitutional procedures.

This distribution reflects the actual frequency of different legal issues raised by Bangladeshi citizens, with family and property law representing the most common concerns.

\subsection{Question Complexity and Characteristics}

The questions in our dataset exhibit varying levels of complexity, which we categorized using a three-tier system:

\textbf{Basic Questions (40\%)}: These involve straightforward factual inquiries about legal procedures, rights, or requirements. Example: ``What documents are needed to file for divorce in Bangladesh?''

\textbf{Intermediate Questions (45\%)}: These require application of legal principles to specific circumstances and may involve multiple legal considerations. Example: ``My husband divorced me through triple talaq, but we have joint property. What are my rights to the property and how can I claim maintenance?''

\textbf{Complex Questions (15\%)}: These involve multiple legal domains, require synthesis of various legal principles, or present novel legal situations. Example: The freedom fighter allowance fraud case shown in our example, which involves criminal law, administrative law, family law, and financial regulations.

This complexity distribution allows us to assess AI performance across different levels of legal reasoning and application.

\subsection{Expert Answer Quality and Verification}

Each question in our dataset includes verified expert answers that serve as our evaluation baseline. These responses were provided by legal professionals whose credentials were verified through the Facebook group's moderation system. Expert answers typically include:

\begin{itemize}
    \item A detailed explanation of applicable law, regulations, and judicial authority that relates to the user's inquiry.
    \item Structured and actionable guidance that clearly addresses the appropriate legal steps, processes, or remedies.
    \item Reasoned assessment of possible challenges or risks and alternatives within the applicable legal context.
    \item Recommendations for any ancillary resources in context, such as legal documents, agencies, contacts, or fur- ther professional help.
\end{itemize}

The expert answers underwent quality review by our research team to ensure they met professional standards and contained sufficient detail for comparative evaluation.

\subsection{Privacy Protection and Ethical Considerations}

All data collection and processing followed strict ethical guidelines to protect user privacy. We implemented several safeguards:

\begin{itemize}
    \item All personally identifiable information, including names, contact details, and addresses were systematically removed or anonymized prior to inclusion in the dataset.
    \item Case-specific details that could inadvertently reveal individual identities or legal proceedings were generalized while preserving their contextual and legal relevance.
    \item Data were collected exclusively from publicly available posts within the Facebook group; no private messages or restricted content were accessed or utilized.
    \item The entire data acquisition protocol underwent institutional ethics review and was approved prior to implementation, ensuring transparency and accountability in research practices.
\end{itemize}

\subsection{Dataset Validation and Quality Assurance}

To ensure dataset quality, we implemented multiple validation steps:

\begin{itemize}
    \item Three independent legal professionals reviewed a representative sample of question-answer pairs to verify the accuracy and appropriateness of expert responses.
    \item Bengali language specialists assessed question clarity to ensure comprehensible and contextually correct phrasing for AI processing.
    \item The dataset was tested with preliminary AI queries to detect formatting or encoding issues that could affect model performance.
\end{itemize}

This comprehensive dataset represents the first large-scale collection of authentic Bengali legal questions paired with verified expert answers, providing a robust foundation for evaluating AI performance in Bangladeshi legal contexts.

\section{Result and Discussion}
\subsection{Automatic Metric-Based Evaluation}
Four large language models (OpenAI 4.1 Mini, Gemini 2.0 Flash, Meta Llama 3–70B, and DeepSeek R1) were used based on an identical input prompt to generate the answers. To evaluate the performance of these models, 50 legal complex questions were selected from the dataset covering wide range of legal domains like  Criminal Law, Inheritance Law,Marriage and Divorce Law, Civil Law,Labor Law etc. For evaluation,  established evaluation metrics including ROUGE, BLEU, METEOR, chrF2, and BERTScore were used comparing the responses against reference answers provided by professional lawyers.

\begin{table}[H]
\centering
\caption{Automatic evaluation metrics for four large language models.\protect\footnotemark}
\label{tab:automatic-metrics}
\small
\renewcommand{\arraystretch}{1.2}
\begin{tabularx}{0.80\textwidth}{l >{\raggedleft\arraybackslash}X >{\raggedleft\arraybackslash}X >{\raggedleft\arraybackslash}X >{\raggedleft\arraybackslash}X >{\raggedleft\arraybackslash}X >{\raggedleft\arraybackslash}X}
\hline
\textbf{Model} & \textbf{R-1} & \textbf{R-2} & \textbf{BLEU} & \textbf{MTR} & \textbf{chrF2} & \textbf{BERT} \\
\hline
OpenAI 4.1 Mini & 0.0011 & 0.0006 & 0.0597 & 0.3283 & \textbf{25.81} & 0.6843 \\
Gemini 2.0 Flash & 0.0000 & 0.0000 & \textbf{0.0674} & \textbf{0.3297} & 25.28 & 0.6882 \\
Meta Llama 3 (70B) & \textbf{0.0033} & \textbf{0.0027} & 0.0647 & 0.3124 & 23.82 & \textbf{0.6896} \\
DeepSeek R1 & 0.0000 & 0.0000 & 0.0615 & 0.3046 & 23.12 & 0.6883 \\
\hline
\end{tabularx}
\end{table}

\footnotetext{\footnotesize R-1 and R-2 refer to ROUGE-1 (F1) and ROUGE-2 (F1) scores, respectively. BERT and MTR refer to BERTScore-F1 and METEOR scores respectively.}

Table \ref{tab:automatic-metrics} presented the automatic metrics evaluation among the four models. There was minimal lexical overlap between the models' answers and the reference answers, as the ROUGE-1 and ROUGE-2 scores were near zero. In terms of the BLEU score, which reflects n-gram similarity, the results were also low, indicating limited n-gram similarity with the reference answers. In contrast, METEOR values were comparatively high, with Gemini 2.0 Flash obtaining the best score of 0.3297. For chrF2, OpenAI 4.1 Mini achieved the highest score of 25.81, while Meta Llama 3 (70B) had the best BERTScore-F1 score of 0.6896.
This discrepency of the results was due to the fact that, LLM-generated answers were more structured and detailed, while the reference responses consisted of informal Facebook comments offering practical guidance and actionable recommendations.

After analyzing the results, it was evident that traditional evaluation metrics captured only minimal aspects of the legal domain. The very low lexical and n-gram overlaps indicated the limitations of these metrics for complex legal reasoning, which motivated the need to adapt a more sophisticated approach by using a state-of-the-art LLM as a judge. This new approach is described in the next section.

\begin{figure}[H]
    \centering
    \includegraphics[width=0.55\textwidth]{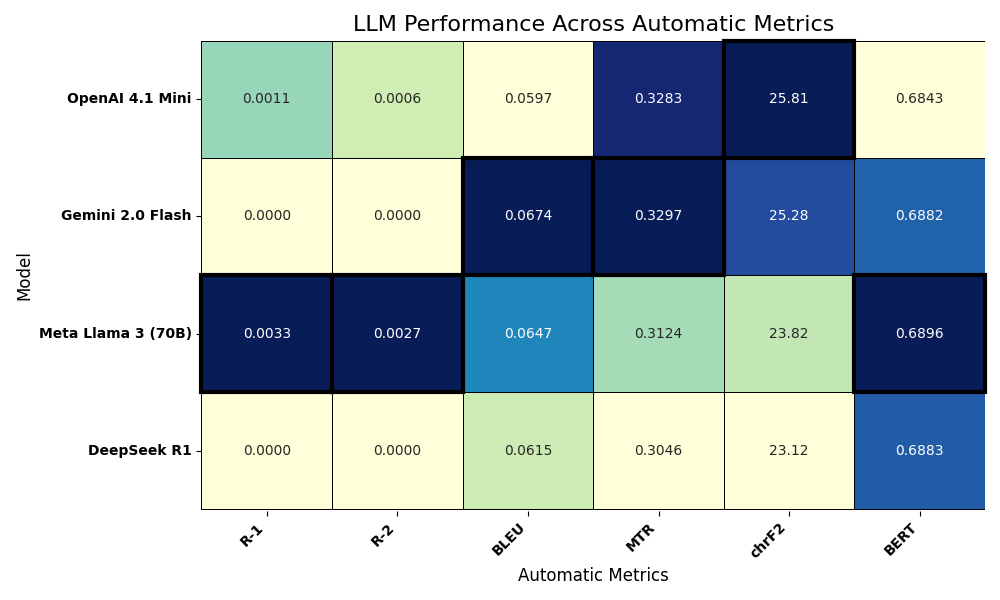}
    \caption{Automatic Metrics Based Evaluation}
    \label{fig:lawyer_spider}
\end{figure}

\subsection{LLM as judge based Evaluation}
As automatic metrics based evaluation was not sufficient enough to capture the complex legal reasoning, a state of the art LLM model was used to evaluate the model performance. Specifically, Gemini 2.5 pro was employed in this purpose.The responses were evaluated based on four specific criteria which were  “Factual Accuracy,” “Completeness,” “Conciseness,” and “Clarity” .

\begin{table}[H]
\centering
\caption{Evaluation of LLM Performance by LLM Judge}
\label{tab:LLM-as-judge}
\small
\renewcommand{\arraystretch}{1.2}
\begin{tabularx}{0.55\textwidth}{l >{\raggedleft\arraybackslash}X}
\hline
\textbf{Model} & \textbf{Average Score} \\
\hline
OpenAI 4.1 Mini & 3.612 \\
Gemini 2.0 Flash & \textbf{3.772} \\
Meta Llama 3 (70B) & 3.177 \\
DeepSeek R1 & 3.266 \\
\hline
\end{tabularx}
\end{table}

Table \ref{tab:LLM-as-judge} showed that among the four models, Gemini 2.0 Flash achieved the highest score of 3.772, suggesting its superior ability to capture complex legal reasoning. This was followed by the next best model, OpenAI 4.1 mini, which achieved an average score of 3.612. In contrast, Meta Llama 3 (70B) was the lowest performing model in this domain, achieving an average score of 3.177.

\begin{figure}[H]
    \centering
    \includegraphics[width=0.45\textwidth]{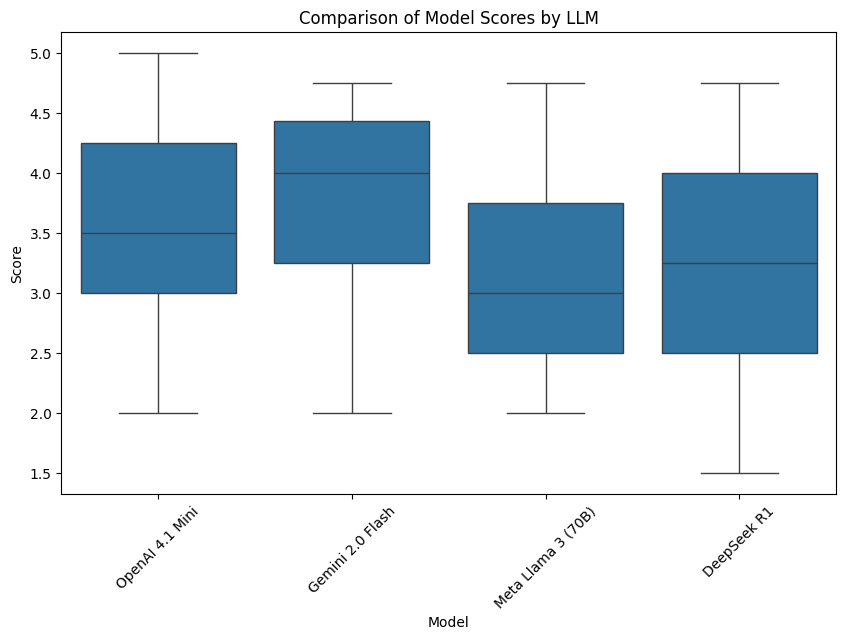}
    \caption{Box plot comparing the distribution of LLM scores based on the LLM-as-a-judge evaluation.}
    \label{fig:anova-boxplot}
\end{figure}
To gain a more comprehensive understanding of the results, and to determine if the differences in average scores are statistically significant, an ANOVA test and a post-hoc Tukey's HSD test were performed.

\begin{table}[H]
    \centering
    \renewcommand{\arraystretch}{1.3}
    \caption{Tukey's HSD Post-Hoc Test Results for Multiple Comparisons of Mean LLM Scores (FWER = 0.05)}
    \label{tab:tukey-hsd}
    \footnotesize
    \resizebox{0.9\textwidth}{!}{%
    \begin{tabular}{|m{0.25\textwidth}|m{0.25\textwidth}|c|c|c|c|c|}
        \hline
        \textbf{Group 1} & \textbf{Group 2} & \textbf{Mean Diff} & \textbf{p-adj} & \textbf{Lower} & \textbf{Upper} & \textbf{Reject $\mathbf{H_0}$} \\
        \hline
        Overall Score by LLM for DeepSeek & Overall Score by LLM for Gemini 2.0 & 0.4956 & 0.0139 & 0.0742 & 0.9171 & \checkmark \\
        \hline
        Overall Score by LLM for DeepSeek & Overall Score by LLM for Meta Llama & -0.0812 & 0.9585 & -0.5005 & 0.3380 & $\times$ \\
        \hline
        Overall Score by LLM for DeepSeek & Overall Score by LLM for Open AI & 0.3450 & 0.1432 & -0.0721 & 0.7621 & $\times$ \\
        \hline
        Overall Score by LLM for Gemini 2.0 & Overall Score by LLM for Meta Llama & -0.5768 & 0.0029 & -1.0004 & -0.1533 & \checkmark \\
        \hline
        Overall Score by LLM for Gemini 2.0 & Overall Score by LLM for Open AI & -0.1506 & 0.7908 & -0.5721 & 0.2708 & $\times$ \\
        \hline
        Overall Score by LLM for Meta Llama & Overall Score by LLM for Open AI & 0.4262 & 0.0447 & 0.0070 & 0.8455 & \checkmark \\
        \hline
    \end{tabular}%
    }
\end{table}

The result of the ANOVA test shows a p-value of 0.0009, indicating a significant difference between the average scores of the models. To identify which specific pairs of models had statistically significant differences, a post hoc Tukey's honestly significant difference (HSD) test was performed.

As shown in Table \ref{tab:tukey-hsd}, the Tukey HSD test revealed three pairs with a statistically significant difference in their mean scores. The first significant difference was found between Gemini 2.0 Flash and Meta Llama 3 (70B) having (p-adj=0.0029). After that, the difference between Gemini 2.0 Flash and DeepSeek R1 was also statistically significant (p-adj=0.0139). Finally, last pair of statistically significant difference observed between Meta Llama 3 (70B) and OpenAI 4.1 (p-adj=0.0447). Apart from these three, there was no statistically significant difference between the models. This result confirms that, Gemini 2.0 Flash is the superior model for the complex bengali legal question answering domain comapred to Meta Llama 3 (70B) and DeepSeek R1. But there was no statistically significant difference found between the top-performing Gemini 2.0 Flash and the next-best OpenAI 4.1 mini indicating Gemini 2.0 Flash's performance is statistically on par with OpenAI 4.1 mini.

\begin{figure}[H]
    \centering
    \includegraphics[width=0.45\textwidth]{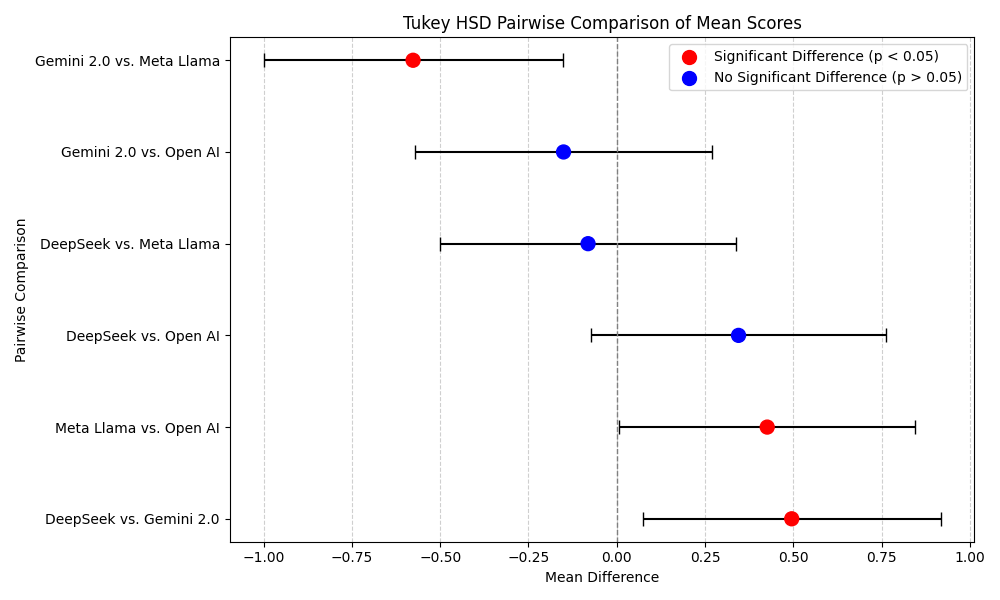}
    \caption{Tukey HSD Pairwise Comparison of Mean Scores based on the LLM-as-a-judge evaluation.}
    \label{fig:tukey-hsd-plot}
\end{figure}

\subsection{Criterion-Based Evaluation by LLM}
To further enhance our reasoning, the models were further compared based on each individual criterion including Factual Accuracy, Completeness, Conciseness, and Clarity to see which model performs better in the specific criteria.

\begin{table}[H]
    \centering
    \caption{Evaluation of LLM Performance based on each criterion by LLM judge}
    \label{tab:LLM-each}
    \footnotesize
    \renewcommand{\arraystretch}{1.3}
    \resizebox{0.65\textwidth}{!}{%
    \begin{tabular}{|l|c|c|c|c|}
        \hline
        \textbf{Model} & \textbf{Factual Accuracy} & \textbf{Completeness} & \textbf{Conciseness} & \textbf{Clarity} \\
        \hline
        OpenAI 4.1 Mini & 3.560 & \textbf{2.940} & 3.460 & 4.500 \\
        \hline
        Gemini 2.0 Flash & \textbf{3.714} & 2.939 & \textbf{3.776} & \textbf{4.653} \\
        \hline
        Meta Llama 3 (70B) & 2.714 & 2.224 & 3.449 & 4.367 \\
        \hline
        DeepSeek R1 & 2.880 & 2.300 & 3.580 & 4.320 \\
        \hline
    \end{tabular}%
    }
\end{table}

As shown in Table \ref{tab:LLM-each}, a consistent trend found across all criteria, with Gemini 2.0 Flash achieving the highest score in each category. For Factual Accuracy as per LLM judge, Gemini 2.0 Flash was able to achieve 3.714 average score out of 5 while the next best model was OpenAI 4.1 Mini with 3.560 average score. In contrast, Meta Llama 3 (70B) and DeepSeek R1 showed significantly lower performance with score of 2.714 and 2.880 respectively.

In case of Completeness, Gemini 2.0 Flash \& OpenAI 4.1 mini achieved nearly identical average score of 2.9 which indicates completeness was challenging criterion in LLM responses.

In terms of Conciseness, Gemini 2.0 Flash showed its superiority by achieving the highest average score of 3.776 while other models achieved nearly same average score.

Finally, in clarity criterion, each LLM showed great results, Gemini 2.0 Flash achieved of highest score 4.653 maintaining its superiority among the models.

\begin{figure}[H]
    \centering
    \includegraphics[width=0.4\textwidth]{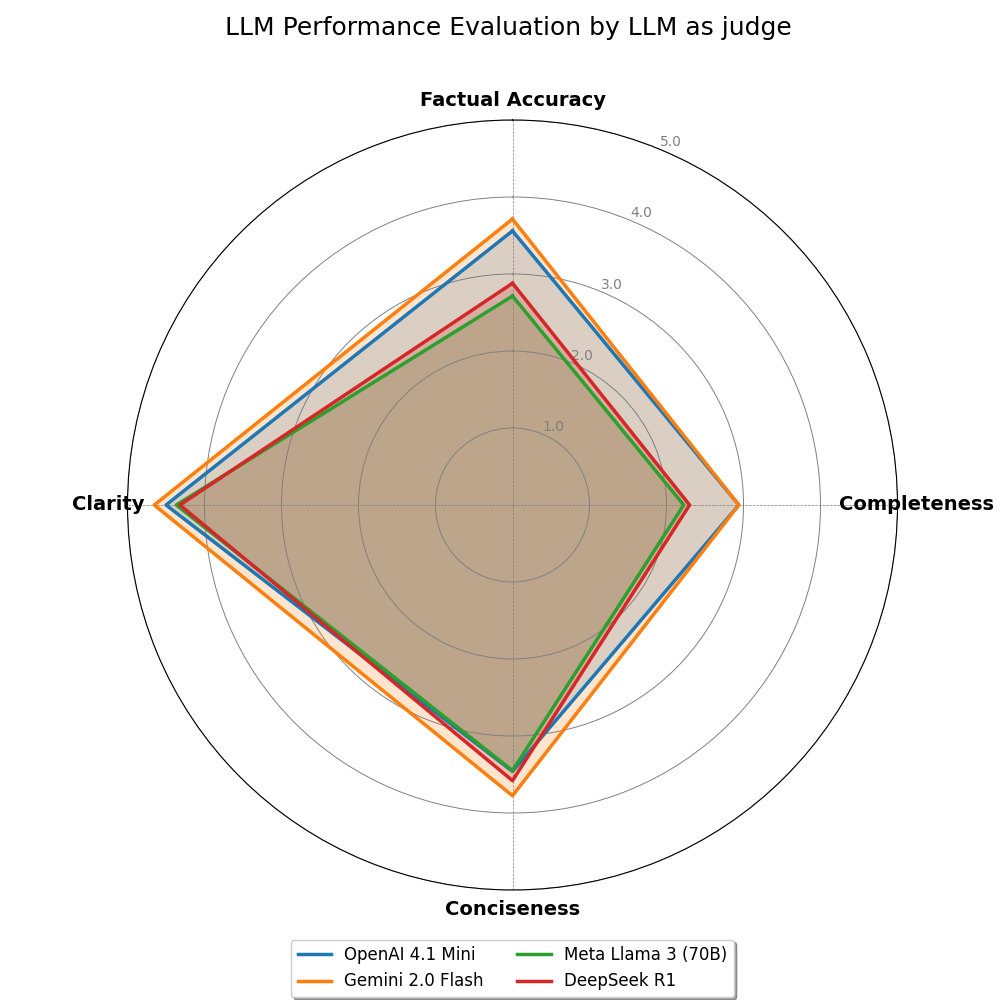}
    \caption{Evaluation of LLM Performance based on each criterion by LLM as judge}
    \label{fig:llm_judge_spider}
\end{figure}

\subsection{Evaluation by professional lawyer}

Though a LLM-as-a-judge provided a more sophisticated evaluation than automatic metrics, its results required validation against real-world human judgment, particularly for complex legal questions. So a hybrid evaluation approach was adopted. which involves using an LLM-as-a-judge to provide an initial assessment and then use professional lawyer evaluation on the same 50 legal questions and their corresponding model responses to validate the reliability of the result against real world human judgment. The hybrid approach of LLM as judge and real world profession lawyer evaluation would provide the ideal standard to find the best model in the domain of legal question-answering.

\begin{table}[H]
\centering
\caption{Evaluation of LLM Performance by Profession Lawyer}
\label{tab:Lawyer-as-judge}
\small
\renewcommand{\arraystretch}{1.2}
\begin{tabularx}{0.6\textwidth}{l >{\raggedleft\arraybackslash}X}
\hline
\textbf{Model} & \textbf{Average Score} \\
\hline
OpenAI 4.1 Mini & \textbf{4.25} \\
Gemini 2.0 Flash & {3.96} \\
Meta Llama 3 (70B) & 3.38 \\
DeepSeek R1 & 3.55 \\
\hline
\end{tabularx}
\end{table}

Table \ref{tab:Lawyer-as-judge} shows that among the four models, OpenAI 4.1 mini achieved the highest score of 4.25 by lawyer making is the best model. Gemini 2.0 Flash which was the top model in the previous LLM as judge evaluation method, was ranked as the second best model by lawyer achieving an average score of 3.96. Additionally, the lawyer based evaluation confirmed that, the Meta Llama 3 (70B) was the lowest-performing model in this domain, achieving an average score of 3.38 a finding that aligned with previous LLM as judge based evaluation.

\begin{figure}[H]
    \centering
    \includegraphics[width=0.45\textwidth]{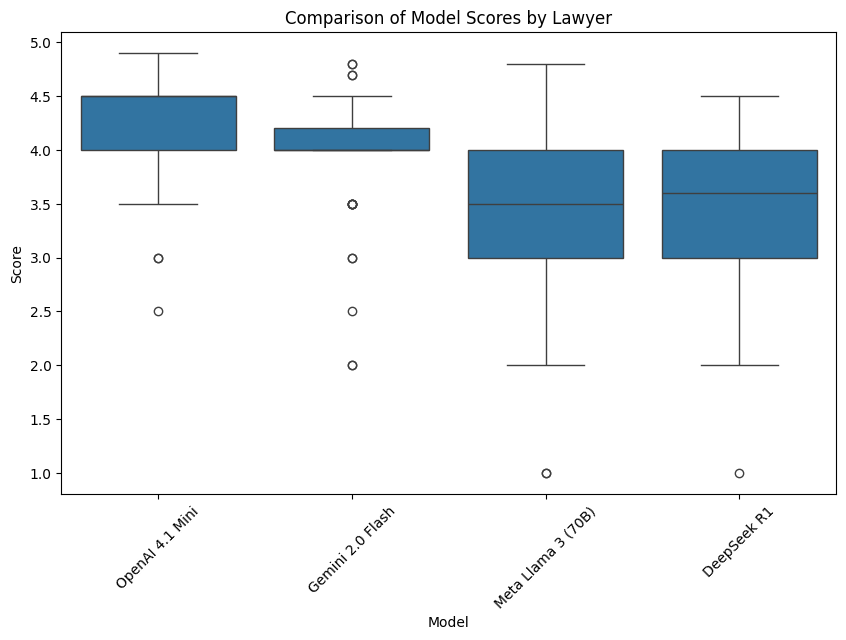}
    \caption{Box plot comparing the distribution of LLM scores based on Professional Lawyer evaluation.}
    \label{fig:lawyer-anova-boxplot}
\end{figure}

As like previous, ANOVA test and a post-hoc Tukey's HSD test were performed again to see check any statistically significant differences in lawyer based evaluation. This statistical analysis was crucial to validate whether the new model ranking, with OpenAI 4.1 Mini as the top performer, was a result of a true difference in performance.

\begin{table}[H]
    \centering
    \caption{Tukey's HSD Post-Hoc Test Results for Lawyer-Based Evaluation (FWER = 0.05)}
    \label{tab:Lawyer-as-judge-tukey}
    \small
    \renewcommand{\arraystretch}{1.3}
    \resizebox{0.95\textwidth}{!}{%
    \begin{tabular}{|m{0.28\textwidth}|m{0.28\textwidth}|c|c|c|c|c|}
        \hline
        \textbf{Group 1} & \textbf{Group 2} & \textbf{Mean Diff} & \textbf{p-adj} & \textbf{Lower} & \textbf{Upper} & \textbf{Reject $\mathbf{H_0}$} \\
        \hline
        Overall Score by Lawyer for DeepSeek & Overall Score by Lawyer for Gemini & 0.398 & 0.0153 & 0.056 & 0.74 & \checkmark \\
        \hline
        Overall Score by Lawyer for DeepSeek & Overall Score by Lawyer for Meta Llama & -0.14 & 0.7138 & -0.482 & 0.202 & $\times$ \\
        \hline
        Overall Score by Lawyer for DeepSeek & Overall Score by Lawyer for Open AI & 0.678 & 0.0000 & 0.336 & 1.02 & \checkmark \\
        \hline
        Overall Score by Lawyer for Gemini & Overall Score by Lawyer for Meta Llama & -0.538 & 0.0004 & -0.88 & -0.196 & \checkmark \\
        \hline
        Overall Score by Lawyer for Gemini & Overall Score by Lawyer for Open AI & 0.28 & 0.1499 & -0.062 & 0.622 & $\times$ \\
        \hline
        Overall Score by Lawyer for Meta Llama & Overall Score by Lawyer for Open AI & 0.818 & 0.0000 & 0.476 & 1.16 & \checkmark \\
        \hline
    \end{tabular}%
    }
\end{table}

The results, presented in Table \ref{tab:Lawyer-as-judge-tukey}, revealed several key findings. A statistically significant difference was found between OpenAI 4.1 Mini and all other models, except Gemini 2.0 Flash. Specifically, OpenAI scored significantly lower than DeepSeek (p-adj $<$ 0.001) and Meta Llama 3 (70B) (p-adj $<$ 0.001), indicating a major difference in performance. However, the comparison between OpenAI 4.1 Mini and Gemini 2.0 Flash was not statistically significant (p-adj = 0.1499). This suggests that while OpenAI 4.1 Mini was rated higher, its performance was statistically on par with Gemini 2.0 Flash.

Additionally, a significant difference was found between Gemini 2.0 Flash and Meta Llama 3 (70B) (p-adj = 0.0004). DeepSeek R1 also differed significantly from Gemini 2.0 Flash (p-adj = 0.0153), but not from Meta Llama 3 (70B) (p-adj = 0.714), suggesting that DeepSeek's performance was intermediate between the two models.

In summary, the lawyer-based evaluation confirms a clear hierarchy among the models, with OpenAI 4.1 Mini and Gemini 2.0 Flash forming a top-tier models whose performance is statistically superior to the other two models.

\begin{figure}[H]
    \centering
    \includegraphics[width=0.45\textwidth]{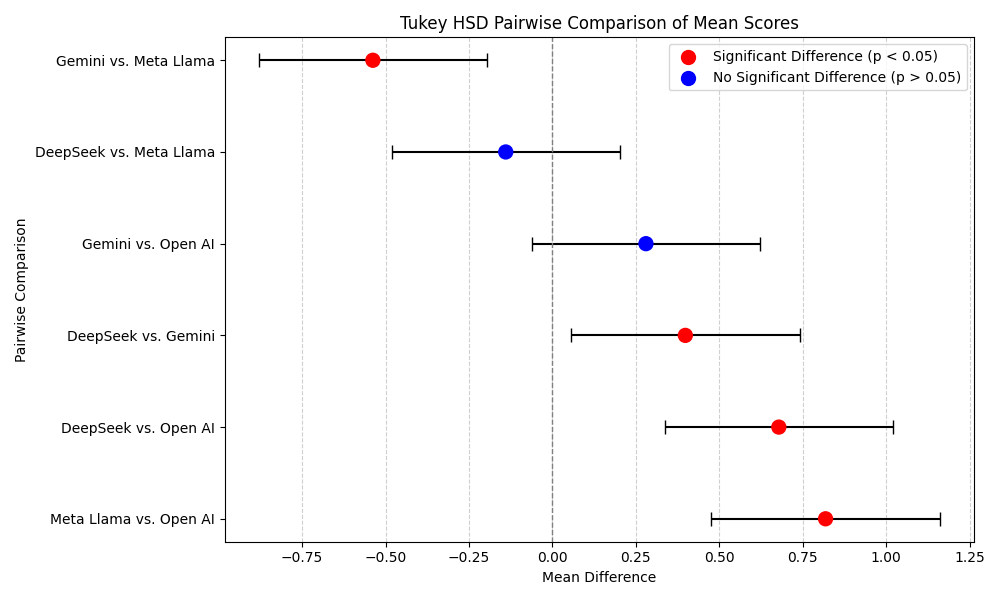}
    \caption{Tukey HSD Pairwise Comparison of Mean Scores based on the Lawyer-as-a-judge evaluation.}
    \label{fig:tukey-hsd-lawyer}
\end{figure}

\subsection{Criterion Based Evaluation by professional lawyer}

\begin{table}[H]
    \centering
    \caption{Evaluation of LLM Performance based on each criterion by Professional Lawyer}
    \renewcommand{\arraystretch}{1.2}
    \label{tab:Lawyer-each-criteria}
    \footnotesize
    \resizebox{0.70\textwidth}{!}{%
    \begin{tabular}{|l|c|c|c|c|}
        \hline
        \textbf{Model} & \textbf{Factual Accuracy} & \textbf{Completeness} & \textbf{Conciseness} & \textbf{Clarity} \\
        \hline
        OpenAI 4.1 Mini & \textbf{4.370} & \textbf{4.330} & 3.930 & \textbf{4.580} \\
        \hline
        Gemini 2.0 Flash & 4.110 & 3.878 & \textbf{4.061} & 4.180 \\
        \hline
        Meta Llama 3 (70B) & 3.490 & 3.194 & 4.050 & 3.660 \\
        \hline
        DeepSeek R1 & 3.590 & 3.450 & 3.850 & 3.850 \\
        \hline
    \end{tabular}%
    }
\end{table}

As like LLM based evaluation, criterion based evaluation was also conducted for Lawyer based evaluation. Unlike LLM based evaluation where Gemini 2.0 Flash was the top performer, OpenAI 4.1 mini emerged as the top in this case.

For Factual Accuracy criterion, OpenAI 4.1 mini was the best model achieving highest score of 4.370 indicating its superior ability to provide correct and verifiable information.

For Completeness criterion it was the best model achieving highest score of 4.330 showing its ability to address more aspects of the query.

In terms of Clarity also OpenAI 4.1 mini was the best model achieving a score of 4.580 indicating its ability to generate responses which is more understandable compared to the other models.

In contrast, Gemini 2.0 Flash achieved highest score in Completeness criterion indicating its responses were more efficient and to the point compared to others.

\begin{figure}[H]
    \centering
    \includegraphics[width=0.45\textwidth]{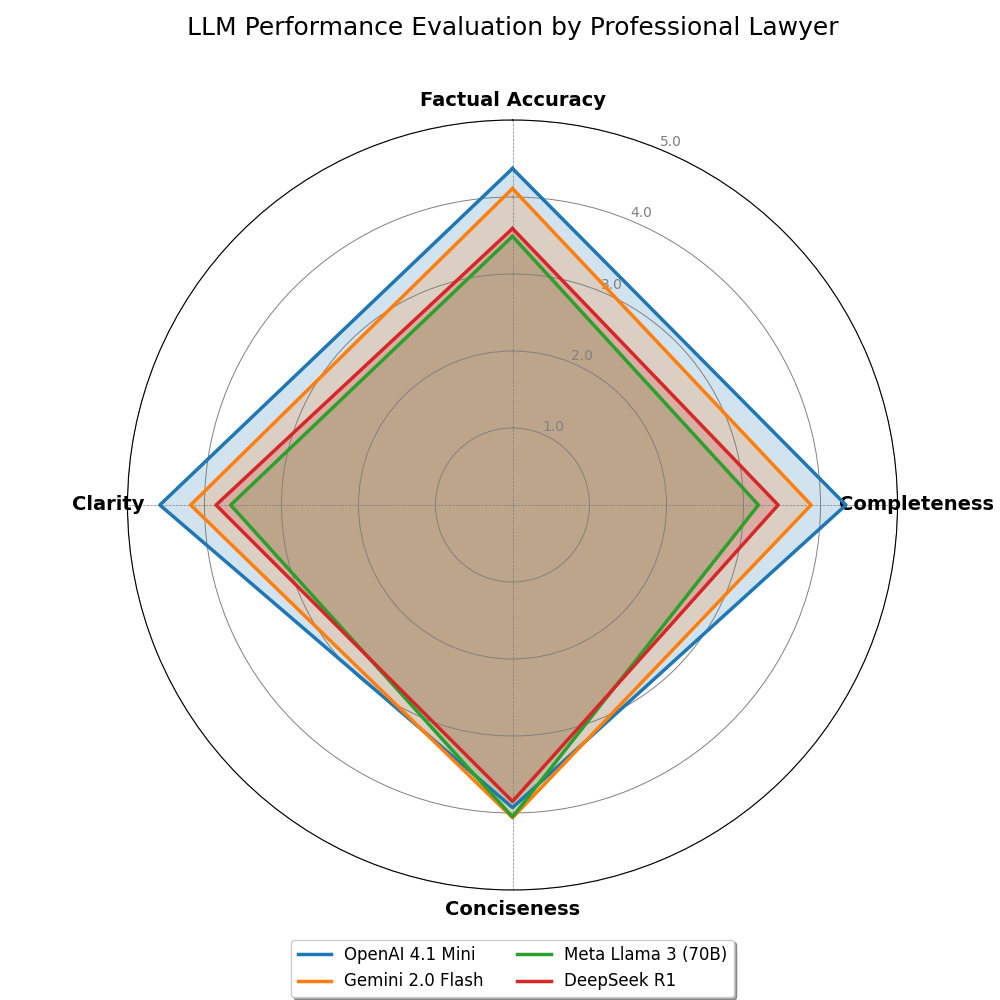}
    \caption{Evaluation of LLM Performance based on each criterion by Professional Lawyer}
    \label{fig:lawyer_spider}
\end{figure}

\subsection{Comparison of LLM and Professional Lawyer Evaluations}

\begin{figure}[H]
    \centering
    \includegraphics[width=0.55\textwidth]{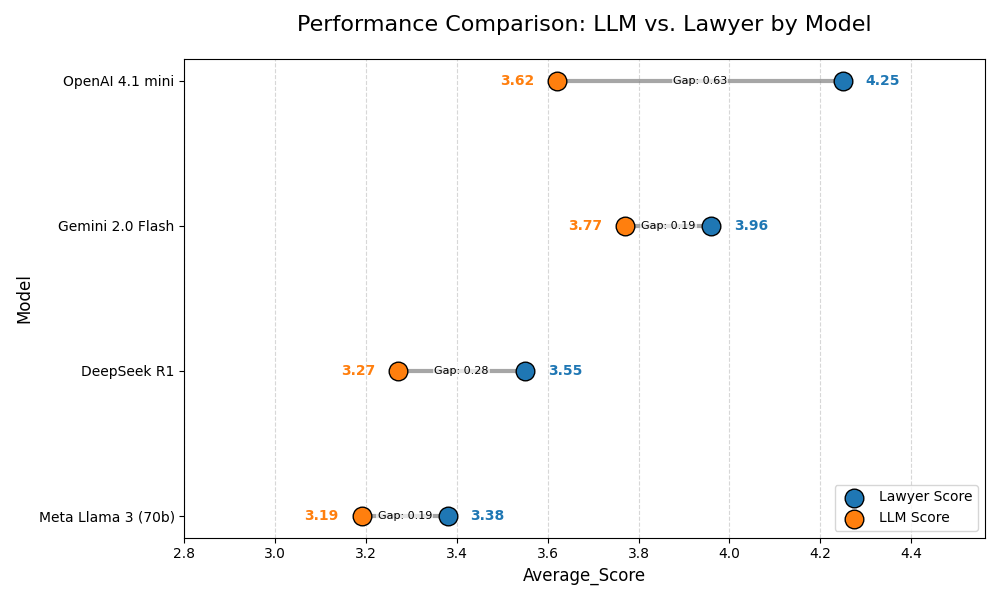}
    \caption*{Figure 5.10 Comparison of Average Scores by LLM-as-a-Judge and Professional Lawyer Evaluations}
    \label{fig:anova-boxplot}
\end{figure}

The evaluation of LLM performance using both an LLM-as-a-judge and professional lawyer revealed a significant discrepancy, highlighting the need for a hybrid evaluation approach. The LLM-as-a-judge initially suggested that Gemini 2.0 Flash was the top-performing model with an average score of 3.772, compared to OpenAI 4.1 mini, which scored 3.612. In contrast, the professional lawyer evaluation produced a different hierarchy where OpenAI 4.1 mini received the highest average score of 4.25, while Gemini 2.0 Flash ranked second with 3.96.

The Tukey's HSD test for the LLM-as-a-judge evaluation showed a p-value of 0.7908 for this pair, and the lawyer-based evaluation showed a similar non-significant p-value of 0.1499 suggesting there was no statistically significant difference between the performance of Gemini 2.0 Flash and OpenAI 4.1 mini in either evaluation and describing both as top tier model.

\begin{table}[H]
    \centering
    \caption{Overall Wilcoxon Signed-Rank Test p-values for OpenAI vs. Gemini Comparison}
    \label{tab:wilcoxon-overall}
    \small
    \begin{tabular}{lc}
        \toprule
        \textbf{Evaluation Method} & \textbf{Wilcoxon \textit{p}-value} \\
        \midrule
        LLM-as-a-Judge & 0.355 \\
        Professional Lawyer & 0.001 \\
        \bottomrule
    \end{tabular}
\end{table}

To go beyond simple score comparisons, a statistical analysis was conducted using the Wilcoxon signed-rank test and Cohen's d to compare the performance of both models. The results showed that a discrepancy between the two evaluation methods. While the lawyer-based evaluation showed a statistically significant overall difference between OpenAI 4.1 mini and Gemini 2.0 Flash (Wilcoxon p-value = 0.001), the LLM-as-a-judge evaluation found no significant difference (Wilcoxon p-value = 0.355).

\begin{table}[H]
\centering
\caption{Effect sizes (Cohen's d) and interpretation for OpenAI vs Gemini (Lawyer-as-judge)}
\label{tab:cohensd}
\small
\renewcommand{\arraystretch}{1.2}
\begin{tabular}{lcc}
\hline
\textbf{Criterion} & \textbf{Cohen's d} & \textbf{Interpretation} \\
\hline
Factual Accuracy   & 0.293  & Small effect; OpenAI slightly better \\
Completeness       & 0.406  & Medium effect; OpenAI moderately better \\
Conciseness        & -0.354 & Small effect; Gemini slightly better \\
Clarity            & 0.804  & Large effect; OpenAI clearly better \\
\hline
\end{tabular}
\end{table}

\begin{table}[H]
\centering
\caption{Effect sizes (Cohen's d) and interpretation for OpenAI vs Gemini (LLM-as-judge)}
\label{tab:cohensd_llm_judge}
\small
\renewcommand{\arraystretch}{1.2}
\begin{tabular}{lcc}
\hline
\textbf{Criterion} & \textbf{Cohen's d} & \textbf{Interpretation} \\
\hline
Factual Accuracy    & -0.091 & Tiny effect; no significant difference \\
Completeness        & 0.000  & No effect; no significant difference \\
Conciseness         & -0.253 & Small effect; Gemini slightly better \\
Clarity             & -0.141 & Small effect; no significant difference \\
\hline
\end{tabular}
\end{table}

As per the result from the table \ref{tab:cohensd} the source of this difference was identified by criterion based analysis. Lawyers rated OpenAI 4.1 mini as significantly superior in Factual Accuracy (p = 0.034), Completeness (p = 0.005), and Clarity (p = 0.000). On contrast, Gemini 2.0 Flash was rated as significantly better in Conciseness (p = 0.036). While from the table \ref{tab:cohensd_llm_judge} there was a very negligible effect on each criterion between the two models indicating that the LLM judge found no meaningful difference between the two models.

From above discussion, it is evident that, though initial findings suggested that, OpenAI 4.1 mini and Gemini 2.0 Flash were the top tier models in this complex legal reasoning domain, but deeper analysis revealed that, between these two, OpenAI 4.1 mini is the superior model. This highlights a notable difference in how human legal expert and the LLM-as-a-judge perceive model performance, suggesting that while OpenAI 4.1 mini is superior according to professional lawyer, the LLM-as-a-judge did not identify a clear advantage for either model.
\section{Limitations}

A significant limitation of this study concerns potential evaluation bias arising from systematic differences in response format between AI-generated answers and authentic expert responses. While evaluators assessed all responses anonymously without knowledge of their source, fundamental differences in presentation style may have inadvertently influenced evaluation outcomes. AI-generated responses consistently demonstrated highly structured formats with detailed legal explanations, systematic statute enumeration, and comprehensive descriptions of legal principles, resembling formal legal memoranda in their thoroughness. In contrast, authentic expert responses originated as informal Facebook comments providing practical guidance and actionable recommendations rather than comprehensive legal education, reflecting practitioners' pragmatic orientation toward real-world problem-solving. This difference may have created systematic bias in evaluation procedures, with evaluators potentially favoring the comprehensive, educationally-oriented AI responses over practically-focused expert responses when assessing completeness and clarity. The detailed explanations provided by AI systems may have appeared more thorough to evaluators, potentially resulting in higher scores despite authentic expert responses being more appropriate for practical legal problem-solving.

Several additional methodological limitations warrant acknowledgment in interpreting these findings. The dataset represents 250 legal questions from a single online platform, potentially limiting generalizability to the broader spectrum of legal issues encountered by Bangladeshi citizens. The six-month collection period may not capture longer-term variations in legal question patterns or seasonal fluctuations in specific legal concerns. The evaluation framework, though comprehensive, relies on subjective expert assessment that may be influenced by individual professional experience and specialization areas, and while inter-rater reliability measures demonstrated substantial agreement, some degree of subjective variation remains inherent in human evaluation of complex legal responses. Furthermore, the focus on Bengali-language questions and Bangladesh-specific legal context, while essential for practical relevance, limits the generalizability of findings to other linguistic and legal contexts where performance patterns may differ substantially due to varying structural characteristics of legal systems or different linguistic properties.

\section{Conclusion}
In this study, we systematically evaluated the performance of four advanced AI models in answering authentic legal questions collected from a verified Facebook group. Using a dual evaluation framework involving both a state-of-the-art LLM model and licensed Bangladeshi legal professionals, we assessed AI responses across factual accuracy, legal appropriateness, completeness, and clarity. Our findings demonstrate that while AI models can provide informative and coherent answers, there remain notable gaps in legal accuracy and context-specific reasoning. As large language models are prone to hallucination, advanced techniques such as Retrieval-Augmented Generation (RAG) can be employed to improve factual reliability. Furthermore, these AI systems have the potential to serve as legal assistants, providing preliminary guidance to the public alongside professional lawyers. Future work could explore model fine-tuning on jurisdiction-specific datasets and the integration of hybrid evaluation strategies to further enhance response reliability

\section*{Acknowledgments}

In the preparation of this manuscript, multiple AI-assisted tools were utilized to enhance the writing process. ChatGPT was employed for paraphrasing, grammatical correction, and language refinement. Grok was used for generating and formatting tables, while Gemini assisted with line formatting and structuring. Additionally, these tools supported tasks such as brainstorming ideas, organizing sections, and analysing feedback for clarity. 
The authors confirm that all content generated or suggested by these
AI tools has been thoroughly reviewed, critically evaluated, and approved by the human authors, who take full responsibility for the accuracy, integrity, and originality of the published work.

We would like to thank ResearchBuddy AI for their support in this research. Research Buddy AI is an AI-powered research collaboration platform that offers services such as supervised lab environments, team-based research work, and opportunities for publication and letters of recommendation through a monthly subscription model.

\bibliographystyle{unsrt}  % or IEEEtran, apalike, etc.
\bibliography{custom}

\end{document}